\documentclass{aa}
\usepackage{graphicx}
\begin{document}
\title{Light curves and H$\alpha$ luminosities as indicators of $^{56}$Ni
 mass in type IIP supernovae}

\titlerunning{$M(^{56}$Ni) and H$\alpha$ in SNe IIP }

  \author{A. Elmhamdi \inst{1} \and N.N. Chugai \inst{2} \and
  I.J. Danziger \inst{3} }

   \offprints{A. Elmhamdi, {elmhamdi@sissa.it}}
   
   \institute{SISSA / ISAS , via Beirut 4 - 34014 Trieste - Italy 
      \and Institute of Astronomy RAS,
  Pyatnitskaya 48, 109017 Moscow, Russia
	\and Osservatorio Astronomico di Trieste, Via G.B.Tiepolo 11 - I-34131
	 Trieste - Italy }
 
\authorrunning{Elmhamdi et al.}

\date{}

\abstract{ 
The possibility is investigated that the H$\alpha$ luminosity at the 
nebular epoch may be an additional indicator of $^{56}$Ni mass in 
type II supernovae with plateau (SNe~IIP), on the basis of available photometry and spectra. We first derive the $^{56}$Ni mass from the
$M_V$ magnitude on the radioactive tail using a standard approach. A confirmation of the correlation between $^{56}$Ni mass and plateau 
$M_V$ magnitude found recently by Hamuy (2003) is evident. There is strong 
evidence of a correlation between steepness of the $V$ light curve 
slope at the inflection time and the $^{56}$Ni mass. If confirmed, 
this relation may provide distance and extinction independent 
estimates of the amount of $^{56}$Ni in SNe~IIP. We then apply 
upgraded radioactive models of H$\alpha$ luminosity at the nebular epoch
and claim that it may be a good indicator of $^{56}$Ni, if mass, energy 
and mixing properties vary moderately (within factor $\sim 1.4$) among SNe~IIP
. This method of the $^{56}$Ni mass determination from H$\alpha$ luminosities 
yields results which are consistent with the photometric mass of 
$^{56}$Ni mass to within 20\%. This result also implies that the parameters 
of SNe~IIP events (mass, energy and mixing properties) are rather similar
among the majority of SNe~IIP, except for rare cases of SN~II intermediate 
between IIP and IIL (linear), of which SN~1970G is an example.
}
\maketitle
\section{Introduction}
Supernovae type II with a plateau in the light curve 
(SNe~IIP) are thought to be a spectroscopically and 
photometrically homogeneous family of core collapse supernovae
(\cite{Filip01}). The light curve of SN~IIP is characterized by 
two distinct 
phases; a plateau with a duration of $60-100$ days and a quasi-exponential 
tail at the later epochs. The plateau corresponds to the radiative 
cooling of the hot opaque envelope originally heated by the 
 explosion (\cite{Grass71}). Variations 
 of observational properties at the plateau phase 
 are interpreted in terms of variation of ejecta mass, explosion 
 energy, and radius of the progenitor (\cite{Litv85}). 
The exponential tail is attributed to the instant reprocessing of the 
 energy of radioactive decay of $^{56}$Co (\cite{Weav80})
 and the luminosity at this stage is directly determined by the 
 mass of ejected $^{56}$Ni, a fact successfully exploited for 
 SN~1987A (\cite{Catch88}).
 
The SN~IIP phenomena are believed to originate from the explosion
 of red supergiant stars whose main sequence masses lie in the range
 $10-25~\rm M_{\odot}$; precise boundary are still a subject of controversy, 
as well as the mechanism(s) of explosion.
The study of optical properties of SNe~IIP provides a means of 
recovering their major characteristics (mass, energy, $^{56}$Ni mass,
asymmetry) and eventually to impose constraints on the 
explosion models and pre-supernova parameters.
The $^{56}$Ni mass is one of the crucial parameters since it 
presumably depends on the presupernova structure and the 
explosion model (\cite{Aufd91}) and can
be directly measured.

The straightforward way to measure $^{56}$Ni mass 
in SNe~IIP is based upon the physical argument that 
the early (120--300 d) bolometric luminosity 
on the radioactive tail should be equal to the luminosity of 
radioactive decay of $^{56}$Co. This is because at this stage the loss of the 
internal energy to expansion ($pdV$ work) is small, while the 
envelope is nearly opaque to gamma-rays.
This method was successfully used for SN~1987A (\cite{Catch88}; \cite{Bouch93}
) with estimates $0.08~\rm M_{\odot}$ and 
0.07~M$_{\odot}$, accordingly, and an average $0.075~\rm M_{\odot}$.
The application of this method to other 
SNe~IIP is, however, hampered by the lack of infrared observations 
longward of the I band. Note, the $UBVRI$ data account for about 
half of the luminosity at this age according to results for 
SN~1987A (\cite{Catch88}; \cite{Schmi93}).
Therefore, for SNe~IIP one uses the absolute flux in one band ({\rm e.g.} $V$) 
 or several  bands in the optical, which, on being compared with the absolute
 flux of SN~1987A provides an estimate of $^{56}$Ni mass.
This method essentially assumes that the  
absolute flux in one or several optical bands
is a constant fraction of the bolometric flux, {\rm i.e.} the 
spectral energy distribution (SED) during the radioactive tail epoch 
is similar for all SNe~IIP. 
Using this approach, Phillips et al. (1990)\nocite{Phill90} claimed that 
similar $B$ and $V$ absolute magnitudes of SN~1987A, SN~1969L and SN~1983K
 on the radioactive tail implies similarity of their ejected $^{56}$Ni masses. 
A similar conclusion was made by Patat et al. (1994)\nocite{Pat94} from a
 study of a large sample of type II SNe. Schmidt et al. (1993)\nocite{Schmi93}
 slightly modified this method applying a procedure of reconstruction of the 
light curve using a bolometric correction determined from SN~1987A data. 
Using a constant bolometric correction 
 does not necessarily improve the accuracy of the determination of $^{56}$Ni
 mass compared with the use of absolute magnitudes in one or several filters. 
Recently Hamuy (2003)\nocite{Ham02} using this method recovered $^{56}$Ni 
mass for an extensive sample of SNe~IIP and pointed to an interesting 
correlation of $^{56}$Ni mass with the luminosity (M$_V$) on the plateau.

Here we revisit the problem of $^{56}$Ni mass in SNe~IIP, pursuing 
two major goals. First, we independently determine the 
$^{56}$Ni mass for a different sample using a unified approach to 
distance determination and present a new correlation between 
$^{56}$Ni mass and light curve shape.
The second major goal is to check the possibility of the 
use of H$\alpha$ luminosity as a tracer of $^{56}$Ni mass.
This possibility is intriguing bearing in mind that 
in the case of high redshift SN~IIP ($z\geq 0.05$) we face a problem of 
contribution of the host galaxy background, which will inhibit the confident
 estimation of the broad band magnitude at the 
radioactive tail phase. On the other hand, H$\alpha$ is not affected 
by the stellar background and thus may be used as an indicator 
of the ejected amount of $^{56}$Ni. Nevertheless incorrect substraction of any
underlying H II region emission could hamper accuracy; identification of this narrow component in the blend would remove this problem.The use of H$\alpha$
 as an indicator 
of $^{56}$Ni mass is prompted by early radioactive models which 
show an expected proportionality between H$\alpha$ luminosity and 
the amount of $^{56}$Ni (\cite{Chug90}). This has also been noted
 in connection with the comparison of light curve and H$\alpha$ 
luminosities for SN~1999em and SN~1987A (\cite{Elmha03}).

The paper is organized as follows. 
We first select a sample of 
well observed SNe~IIP with late photometry and spectra, 
and  adopt appropriate distances and extinction (Sect. 2).
We then apply the $V$ light curve technique 
to estimate the $^{56}$Ni mass for these
supernovae and study correlations between 
the  $^{56}$Ni mass
and light curve shape (Sect. 3). Then we use
the late H$\alpha$ luminosity to derive $^{56}$Ni 
mass (Sect. 4). Both photometric and specroscopic $^{56}$Ni mass 
are then compared and the implications are discussed (Sect. 5).
\section{Sample, distance, extinction}
\begin{figure}
\includegraphics[height=10cm,width=9cm]{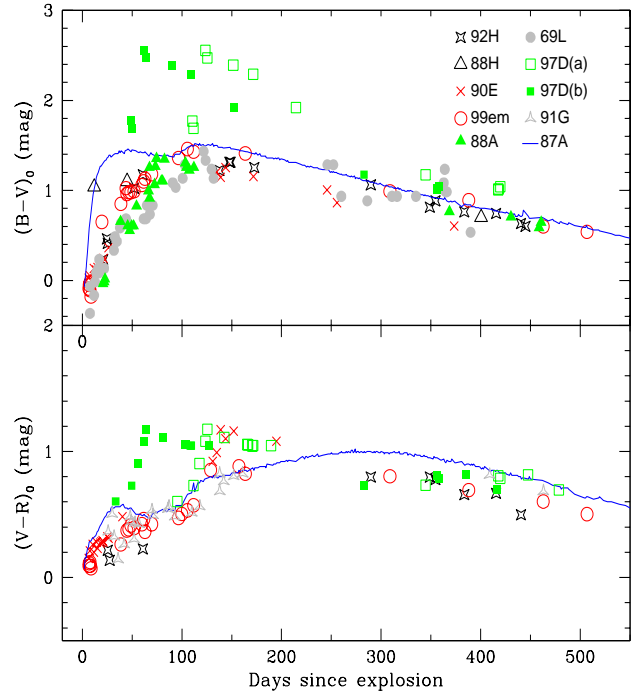}
\caption{ The  $B-V$ (upper panel) and  $V-R$ (lower panel) intrinsic colour
 evolution of the SNe sample. Both options for SN 1997D are shown (sect. 4.2). The used reddening is reported in Table 1 (fourth column). }
\label{colour} 
\end{figure}
Our basic sample consists of 9 SNe IIP, namely: SN 1987A, SN 1969L, SN 1988A, SN 1988H, SN 1990E, SN 1991G, SN 1992H, SN 1997D and SN 1999em, although throughout the paper we will be invoking other 
 objects (SN 1970G, SN 1995ad, SN 1995V, SN 1995W, 
 SN 1999gi and SN 1999eu). The events are selected on the basis of published and unpublished photometric and spectroscopic observations at the plateau and later epochs. The sample parameters are presented in Table 1 together with the corresponding references to the literature from which the 
photometry has been taken. In what follows we adopt the standard 
reddening laws of Cardelli et al. (1989)\nocite{Card89}.
\begin{table*}
\begin{minipage}{100mm}
 \caption{Parameters data of the SNe IIP sample} 
\bigskip
\begin{tabular}{c c c c c c }
\hline \hline
SN &Parent  &Distance & $A_{V}^{tot}$ & $M_{V}$ late slope & References   \\
name &galaxy & (Mpc) &  &mag (100d)$^{-1}$ & \\
\hline
1987A&LMC&0.05 &0.6&0.976 & 1, 2\\
1999em&NGC 1637&8.8&0.31&0.97&3, 4, 5  \\
1969L&NGC 1058&9.057&0.203&0.97 &6, 7 \\
1988A&NGC 4579&22.95 &0.136&1.12 &8, 9 \\
1988H&NGC 5878&28.47 &0.47&$---$ &9 \\
1990E&NGC 1035&16.18 &1.2 &0.86 &11, 14 \\
1991G&NGC 4088 &14.11 &0.065 &1.11 &10 \\
1992H&NGC 5377&29.07 &0.054 & 0.97 &15 \\
1997D&NGC 1536 &16.84 &0.07 &0.87 &12, 13, 16 \\
1970G&NGC 5457 &7.2&0.4&1.04&6, 17 \\
1999gi&NGC 3184 & 10.91 & 0.65 &$---$& 18 \\ 
1999eu & NGC 1097 & $---$&$---$&$---$& 19 \\
\hline \hline 
\end{tabular} 
\end{minipage}\\
\emph{{\rm \scriptsize REF:\\1- Arnett 1996\nocite{Art96}; 2- Hirata et al. 1987\nocite{Hir87}; 
3- Baron et al. 2000\nocite{Bar00}; 4- Elmhamdi et al. 2003;\\ 5- Hamuy et al. 2001\nocite{Ham01}; 6- Kirshner \& Kwan 1975; 7- Patat et al. 1994; 8- Ruiz-lapuente et al. 1990\nocite{Ruiz90}; 
\\9- Turatto et al. 1993\nocite{Tur93};
10- Blanton et al. 1995\nocite{Blan95}; 11- Benetti et al. 1994\nocite{Ben94}; 
12- Turatto et al. 1998;\\ 13- Benetti et al. 2001\nocite{Ben01}; 14- Schmidt et al. 1993;
 15- Clocchiatti et al. 1996\nocite{Clocc96}; 16- Zampieri et al. 2002 \\ 17- Barbon et al. 1973; 18- Leonard et al. 2002; 19- Pastorello et al. 2003\\
$\ast$ H$_{0}$=70 km s$^{-1}$Mpc$^{-1}$ is adopted. }}
\normalsize
\end{table*}
For the problem of $^{56}$Ni determination the total 
extinction and distance estimates are of principal importance.
Galactic extinction is removed using the map
of galactic dust extinction by Schlegel et al (1998)\nocite{Schl98}. The host
 galaxy  reddening is then estimated from the $B-V$ and $V-R$ colour excess 
compared to the intrinsic colour curves of SN 1987A. 
This approach is justified by the fact that at the late photospheric 
phase SNe IIP seem to
follow colour evolution similar to SN 1987A (\cite{Schm92}). 
The use of the $B-V$ colour 
alone to estimate the host galaxy reddening may present some problems, 
as in most cases it provides negative reddening. This point was noted 
also by Hamuy (2003)\nocite{Ham02} who used the $V-I$ index for an independent
 estimate of extinction. 

The recovered total visual reddening ($A_{V}^{tot}$) of our 
sample is listed in Table 1. Fig. 1, on the other hand, displays the 
intrinsic $B-V$ and $V-R$ colour evolution. Note that for the case of 
SN 1997D, both colours show a large excess by the end of the 
photospheric phase. The small (case b; \cite{Tura98}) and large (case a; \cite{Zamp02}) age scenarios are displayed. 
This high reddening seems unlikely since the presence of 
interstellar lines was not reported in the early spectra as one may expect 
in a case with such high reddening (\cite{Tura98}). The very red 
colour at the end of the photospheric phase is seen in other objects 
with a very low ejected $^{56}$Ni mass ({\rm e.g.} SN 1999eu; Pastorello et al. 2003, in preparation). An explanation of this peculiarity may be related to 
the fast cooling from the plateau phase to reach the faint 
radioactive tail, and/or possibly relates to the nature and structure 
of the progenitor star.
Interestingly, the late nebular intrinsic $B-V$ colours of the complete sample seem to be very 
similar to that of SN 1987A, 
 both in slope and magnitude. The convergence 
to the same fate at later phases is also confirmed for the faint object 
SN 1997D and as well as for the underluminous SN 1999eu 
(Pastorello et al. 2003). 
Although studying the spectral energy distribution 
(SED) of SNe IIP is beyond the scope of the present work,
 the fact of a small scatter ($\sim \pm 0.3$ mag) 
 in the sample around the SN 1987A intrinsic $(B-V)_{0}$ 
 colour indicates the similarity 
 in the SED for SNe IIP on the radioactive tail.   
\begin{figure}
\includegraphics[height=9.5cm,width=9cm]{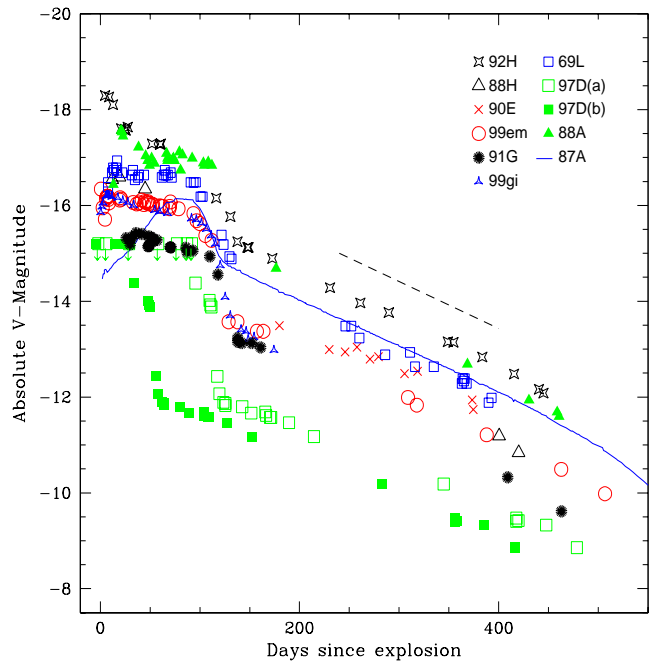}
\caption{ The absolute $V$-magnitude evolution of the SNe sample. 
The used parameters are reported in Table 1. Both possibilities for SN 1997D are plotted. The dashed line shows the slope for $^{56}$Co decay.}
\label{lcurve}
\end{figure}                                      
The distance of $\rm D= 50$ kpc is used for SN 1987A,
 while for the other SNe we adopt distances derived from 
 the recession velocity of the host galaxy  
 corrected for Local Group infall onto the Virgo Cluster as reported 
in the 
``LEDA\footnote{available online at http://leda.univ-lyon1.fr/cgi-bin/single.pl}''
 extragalactic database. 
The Hubble constant $\rm H_{0}$=70 km s$^{-1}$Mpc$^{-1}$ is adopted. 
 The computed distances are reported in third column of Table 1. We note here
that a possibly more accurate value of the distance to SN 1999em, 7.8 Mpc,
 comes from stellar studies of the parent galaxy NGC 1637 (\cite{Sohn98}).
\section{$^{56}$Ni mass from $V$ light curve}
\begin{table}
\begin{minipage}{80mm}
 \caption{$^{56}$Ni mass estimates from photometry} 
\bigskip
\centering
\begin{tabular}{cccc}
\hline \hline
SN  & $M_{\rm Ni}(V)$& t$_{\rm i}$ \\
name &$(M_{\odot})$ & (days) \\
\hline
1987A&0.075 &107 ($\pm2$) \\
1969L &0.067 ($\pm0.002$)&110 ($\pm4$) \\
1988A &0.088 ($\pm0.003$)& 134 ($\pm4$) \\
1988H &0.033 ($\pm0.004$)& $---$ \\
1990E &0.043 ($\pm0.0024$)&$---$ \\
1991G &0.021 ($\pm0.0032$)& 122 ($\pm5$) \\
1992H &0.123 ($\pm0.002$)&112 ($\pm5$) \\
1997D(a) &0.0065 ($\pm0.0003$)&112 ($\pm5$) \\
1997D(b) &0.0036 ($\pm0.0002$)& 112 ($\pm5$) \\
1999em&0.027 ($\pm0.002$)&112 ($\pm4$) \\
1970G & 0.051 ($\pm0.003$)&94 ($\pm4$) \\
1999gi& 0.0246 ($\pm0.0007$)& 120 ($\pm3$) \\
1999eu& 0.0028$^\ast$&110 ($\pm4$) \\
\hline \hline
\end{tabular}
\end{minipage}\\
\\
\centering{\emph{{\rm \scriptsize
$^\ast$ From Pastorello et al. 2003 (In preparation)}}}\\ 
\hspace{-7truecm}\emph{\rm }
\normalsize
\end{table}

The luminosity of SN~IIP on the radioactive tail is 
 controlled by the radioactive decay 
 ($^{56}$Co$\rightarrow$$^{56}$Fe) and if the trapping of 
 gamma rays is efficient, the decline rates of the light
  curves of SNe II should converge to the exponential 
  life-time of $^{56}$Co ({\rm i.e.} 111.26 days or 0.976 mag per 100 d). 
  The radioactive tails in SN~IIP all show similar decay rates, especially in 
  the $V$ band (\cite{Pat94}). This fact provides a solid basis 
  for using the $V$ light curve for the  
  recovery of the ejected $^{56}$Ni mass 
  (Phillips et al. 1990; Schmidt et al. 1993). In Fig. 2 we display
 the absolute $V$ light curves of the SNe sample
 together with that of SN 1987A. The computed late time decline slopes 
 ($150-400$ days since explosion) are reported as well in Table 1 
 (fifth column). Their mean value is 
 $<\gamma^{V} >~ \simeq ~0.99~ (\pm 0.13)$, consistent with 
 the radioactive decay of $^{56}$Co and consequent trapping 
 of the gamma-rays.
  
\begin{figure*}
\center
\includegraphics[height=11.5cm,width=17.5cm]{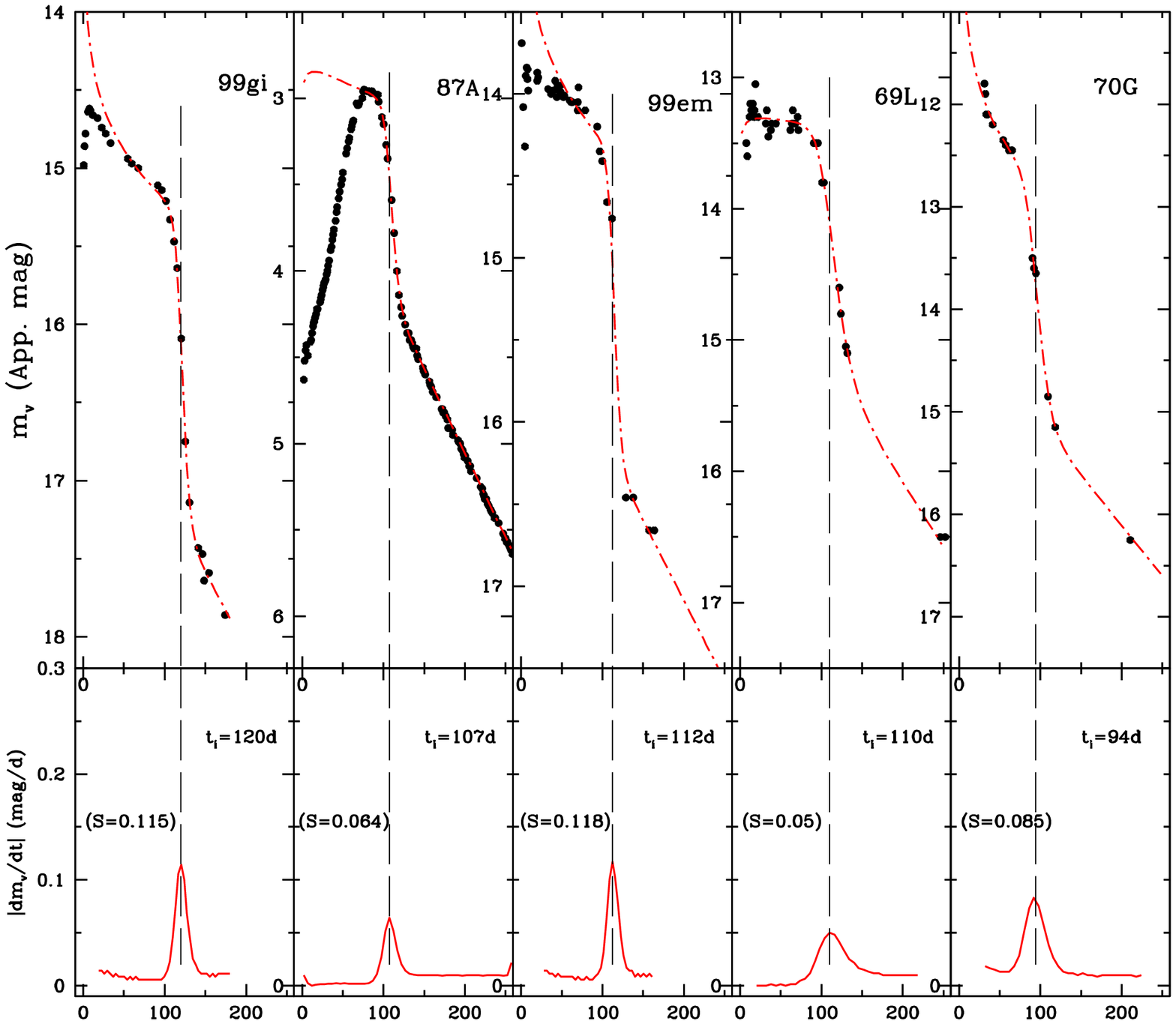}
\vspace*{-1truecm}
\center
\includegraphics[height=11.5cm,width=14cm]{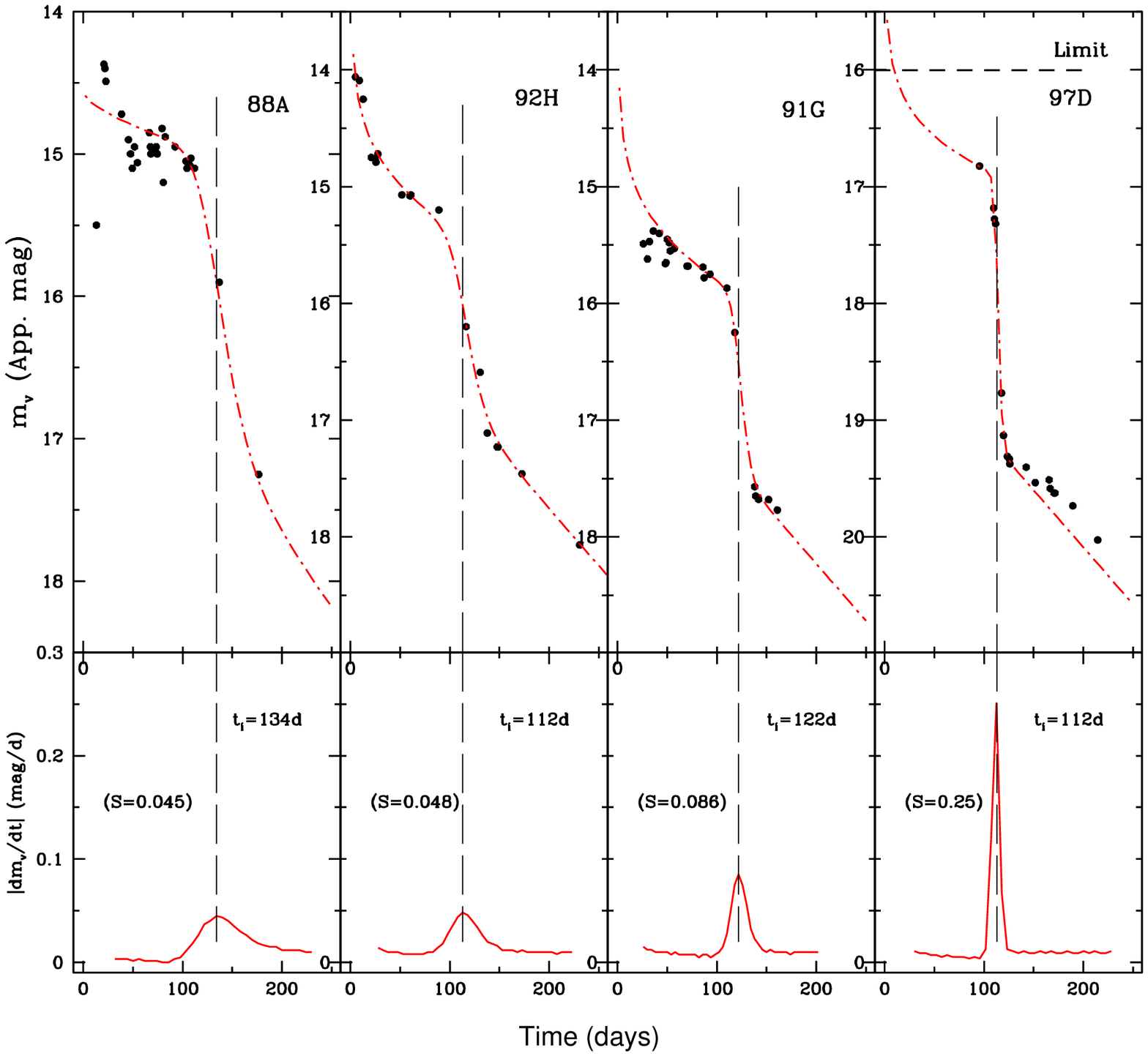}
\caption{  Determination of the steepness and moment of inflection. 
For each SN of the sample: 
upper panel displays the $V$ light curve (dotted points) together
  with the best fit (dashed curve) while lower panel shows the steepness 
   $S$ (see text). The inflection point $t_{\rm i}$ 
  shown by dashed line corresponds to the maximum of $S$. SN 1999eu is not shown because the data is not yet published. 
 } 
\label{inflec} 
\end{figure*} 
The $^{56}$Ni mass is estimated from absolute $M_V$ magnitudes
  between $120-400$ d using the SN 1987A tail as a template.
Results produced by the least squares fit 
are reported in Table 2. Both possible cases for SN 1997D are reported (more 
details concerning this event are discussed in sect. 4.2). 
 These values show a significant range of the 
ejected $^{56}$Ni masses, from $0.0028~\rm M_{\odot}$ and $0.0036~\rm M_{\odot}$ for 
subluminous SN 1999eu and SN 1997D(case b) respectively to $0.123~\rm M_{\odot}$ for SN 1992H with the 
average value 
$\approx 0.05~\rm M_{\odot}$. Our values 
differ somewhat from those determined recently by Hamuy (2003) for objects in 
common. The differences stem primarily from the different distances adopted 
here. As noted earlier the methodology followed by Hamuy is based upon the 
conversion of $V$ magnitude into a bolometric luminosity through an assumed 
constant bolometric correction, while we are adopting SN 1987A luminosities 
as a template at late epochs.  

With our values of the $^{56}$Ni masses it is instructive to
 check that the correlation between plateau $M_V$ magnitude and the amount
 of $^{56}$Ni found by Hamuy (2003) is preserved. 
We have modified the definition of the plateau $M_V$ magnitude 
 compared to Hamuy (2003) in order to avoid any 
uncertainty in the explosion time. Therefore, we use the inflection time
during transition from plateau to the radioactive tail as 
 a zero point. The inflection time $t_{\rm i}$
is defined as the moment when the first derivative at the 
 transition phase $S=-dM_V/dt$  (we dub it ``steepness")
 is maximal.
To calculate $S$ we use the following procedure.
The $V$ band flux in the transition period from plateau and 
radioactive tail is approximated by a sum of plateau and radioactive terms:

\begin{equation}
 F= A\frac {(t/t_{0})^{p}}{1 + (t/t_{0})^{q}}+B\exp (-t/111.26),    
\end{equation}

\noindent where $A$, $B$, $t_{0}$, $p$ and $q$ are parameters derived by 
the $\chi ^{2}$ minimization technique in the sensitive 
interval $t_{\rm i}\pm50$ days. The behaviour of $S$ and determination of $t_{\rm i}$ is demonstrated for
 each SN in Fig. \ref{inflec}. The derived $t_{\rm i}$ are reported in Table 2 (third column) with the corresponding errors while the computed $S$ values and their errors are presented in Fig. 5. The errors in $S$ and $t_{\rm i}$ are estimated according to the best fit (shown in Fig. 3) and as well by introducing test points for events with poor data especially at the transition phase (from plateau to radioactive decay).
With the determined inflection time $t_{\rm i}$ we choose the 
epoch $t_{\rm i}-35$ days as a reference for the $M_V$ on the plateau.

\begin{figure}
\includegraphics[height=10cm,width=9cm]{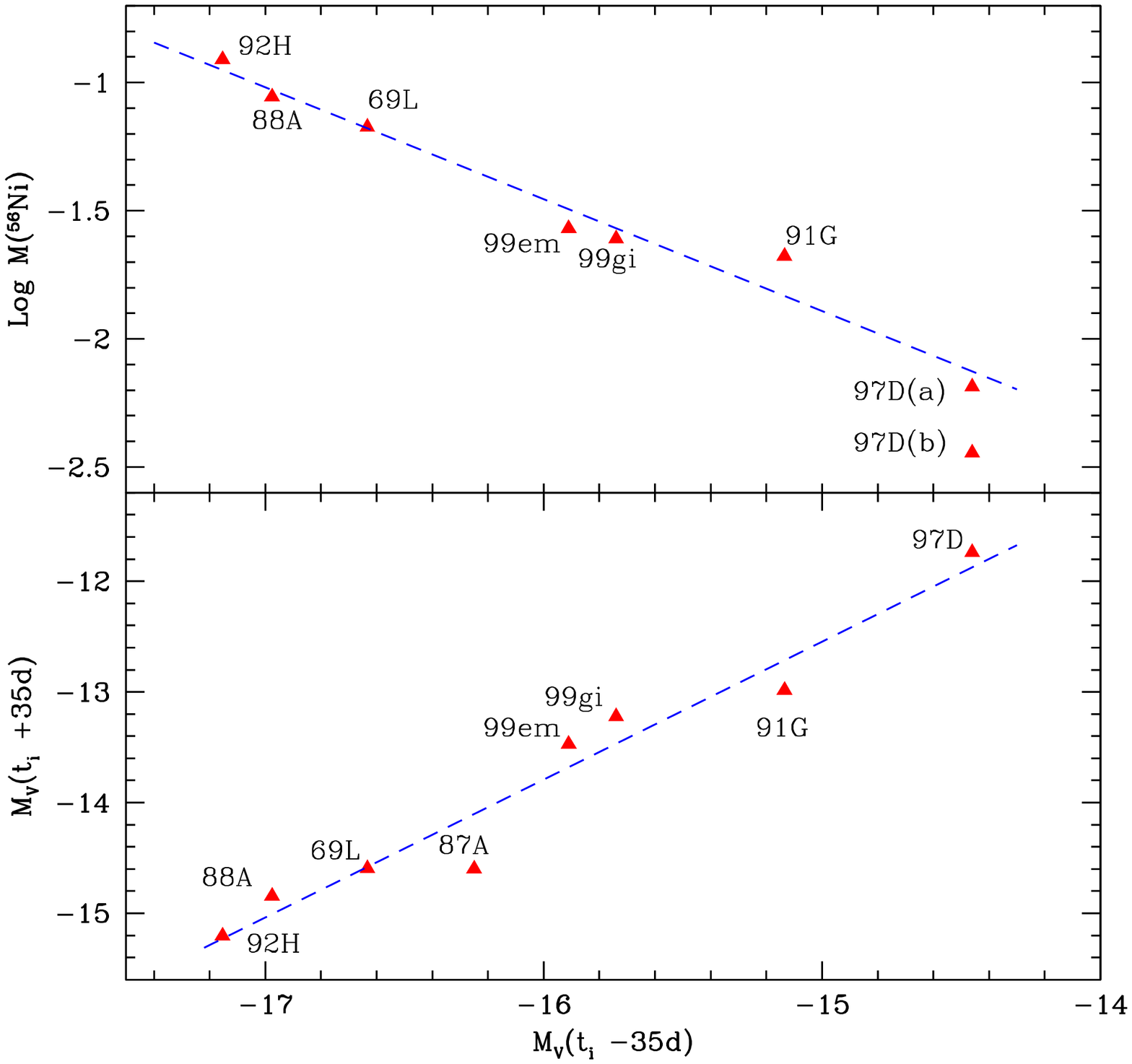}
\caption{The correlation between $M_V$ magnitudes of plateau and 
of radioactive tail. Upper panel shows the correlation between 
 $M_V$ at the moment ($t_{\rm i}-35$d) and $^{56}$Ni mass derived 
 from the tail magnitudes; case ``a'' of SN 1997D is adopted for the fit. Lower panel shows directly the 
 correlation of plateau magnitude $M_V(t_{\rm i}-35{\rm d})$ and tail 
 magnitude $M_V(t_{\rm i}+35{\rm d})$.
 }
\label{mvni}
\end{figure}

Fig.~\ref{mvni} (upper panel) demonstrates the correlation between 
photometric $^{56}$Ni mass and the absolute magnitude $M_V(t_{\rm i}-35)$ 
for the sample of SNe~IIP with measurable $S$ and  $t_{\rm i}$ values.
This plot shows small scatter around a linear trend and thus 
supports the correlation found by Hamuy (2003). In our case, and assuming 
case ``a'' (Zampieri et al. 2002) for SN 1997D, the 
linear correlation is described by the equation:

\begin{equation}
\log\,M(^{56}{\rm Ni})=-0.438M_V(t_{\rm i}-35)-8.46. 
\end{equation}
\noindent Therefore this result suggests a direct correlation between 
$V$ magnitudes (absolute or apparent) of plateau and radioactive tail. In fact 
Fig.~\ref{mvni} (lower panel) demonstrates the correlation between 
 $M_V$ magnitudes at plateau (on $t_{\rm i}-35$ day) and tail 
 (on $t_{\rm i}+35$ day).This is  therefore a direct form of 
 the correlation shown in the upper panel.

\begin{figure}
\includegraphics[height=9.5cm,width=9cm]{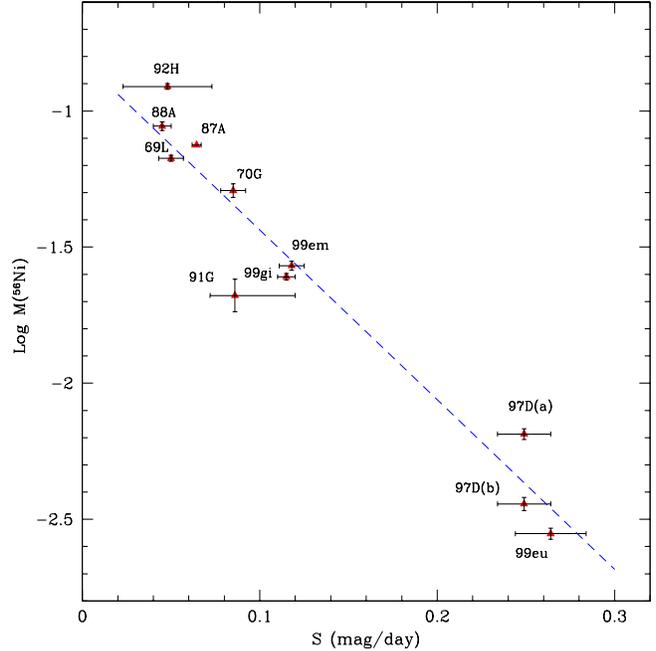}
\caption{The correlation between $^{56}$Ni mass and steepness $S$. Case ``a'' of SN 1997D is adopted for the linear fit (dashed line).} 
\label{smni}
\end{figure}

The sample we explore reveals another interesting correlation.
Fig.~\ref{smni} demonstrates that the steepness $S$ anticorrelates 
with $^{56}$Ni mass:
the lower is the $^{56}$Ni mass the larger is $S$, {\rm i.e.} the steeper is 
the transition from plateau to the tail. 
We note however, that an accurate determination 
of $S$ necessitates a $V$ light curve with a reasonably high density of
 observational points at the end of the plateau 
phase and the beginning of the radioactive tail. Indeed 
SNe 1990E and 1988H are not studied here because of the paucity 
of their light curve data. 
The best linear fit (Fig.~\ref{smni}) reads: 

\begin{equation}
\log\,M( ^{56}{\rm Ni}) = -6.2295\,S -0.8147
\end{equation}

\noindent The interpretation of this correlation requires
hydrodynamical modeling with different amounts of
$^{56}$Ni and degrees of mixing.
In the case of unmixed $^{56}$Ni one does not expect a notable
dependence of the steepness on the amount of $^{56}$Ni. This
is readily seen from the modeling SNe~IIP light curves
by Eastman et al. (1994)\nocite{East94}. For the same amount of $^{56}$Ni
the mixing results in a decreased steepness because of an
increase of radiative diffusion (\cite{East94}). Unfortunately,
in the cited paper the authors did not model cases of variable $^{56}$Ni
mass with similar degrees of mixing, so the question of the
physics behind the above correlation, remains open.
At this stage, we may only suggest that somehow the increase
of the $^{56}$Ni mass in SNe~IIP ejecta favours the larger contribution
of radiative diffusion at the end of the plateau and, therefore, a
less steep transition from plateau to the radioactive tail. It may well be
 that the increase of the $^{56}$Ni mass is accompanied by the growth
 of the degree of mixing degree which favours a less steep decline as 
demonstrated in Fig. 10 and Fig. 12 of Eastman et al. (1994). 

Interestingly, if the correlation between steepness and
$^{56}$Ni mass is confirmed, this will provide us with an
exciting possibility of probing $^{56}$Ni mass in SNe~IIP
 from the light curve shape in a manner independent of
 the distance and extinction.
%
\section{$^{56}$Ni mass from H$\alpha$ luminosity}
The H$\alpha$ luminosity of SNe~IIP at late epochs reflects 
the total ionization in the weakly ionized gas 
caused by radioactive decay of $^{56}$Co 
and, therefore, to a first approximation 
 is proportional to the overall deposition rate and hence 
 to the $^{56}$Ni mass  
(\cite{Chug90}; \cite{Xu92}; \cite{Koz92}) 
unless effects due to higher densities become important. 
This suggests that the H$\alpha$ luminosity resulting from 
 gamma-ray trapping may provide a quantitative measure of
 the $^{56}$Ni mass in SNe~IIP. The accurate measurements on 
SN~1987A provide a control. To explore deviations from the simple
 proportionality we use an upgraded radioactive model of the
 H$\alpha$ luminosity.
\subsection{Model of H$\alpha$ luminosity}
\begin{table*}
\caption{Parameters of the H$\alpha$ evolutionary models}
\bigskip
\begin{center}
\begin{tabular}{lcccccccccccc}
\hline\hline
Parameter& Unit   &  Standard  &   a & a1  & b & b1 & b2 & 
    c      &      d    &     e     &  f & f1 \\
\hline
$M$& $M_{\odot}$  &   14       &    10   & 7 &  14 &14&14  & 
   14      &      14   &    14     &  14 & 14\\
$E$ &$10^{51}$    &    1       &    1   & 1 & 1 &1&1   &
  1.5      &       1   &    1      &   1 & 1\\
$M_{\rm Ni}$ & $M_{\odot}$  &   0.075    & 0.075 & 0.075  & 0.0375 & 
 0.00375&0.15 &0.075     &   0.075   &  0.075    &  0.075 & 0.075\\
$f_{\rm mix}$ &             &    0.4     &   0.4  & 0.4 & 0.4&0.4 &0.4   & 
  0.4      &     0.4   &   0.1     &   0.4 & 0.4\\
$f_{\rm met}$  &         &   0.2      &    0.2 & 0.2 & 0.2 &0.2 &0.2  & 
    0.2    &    0.2    &   0.2    &  0.3 & 0.6\\
$T_{\rm c}$ &K    &  7000      &  7000  & 7000 & 7000 &7000&7000&
  7000    &   5000     &  7000    & 7000 & 7000\\
\hline\hline 
\end{tabular}
\label{tmod} 
\end{center}
\end{table*}
Some brief comments about the modified model 
of H$\alpha$ luminosity powered by the radioactive decay are in order.
The primary purpose of the modification of the previous version
(Chugai 1990) is to specify better the early nebular phase.
This was done by the implementation of macroscopic mixing, collisional 
and radiative de-excitation effects. The supernova envelope is mimicked 
by two zones. The inner zone
with the radius $r_{\rm mix}=v_{\rm mix}t$ is
 a mixing core with a total mass fraction $f_{\rm mix}$,
which is divided between metals together with He-rich matter 
(we dub this mixture ``metals" with the mass fraction of $f_{\rm m}$) and 
H-rich matter with the mass fraction  $1-f_{\rm mix}$. 
The outer zone consists  of hydrogen-rich matter.
All the $^{56}$Ni resides in the mixing core and has a generally 
clumpy distribution. This is mimicked by 
 $^{56}$Ni clumps in cocoons of metals with the 
 total area of cocoons being 
 $4\pi r_{\rm mix}^2 f_{\rm s}$. Here $f_{\rm s}$
is a mixing parameter, which is unity if all metals reside 
in the  spherical 
layer with the inner spherical cavity occupied by $^{56}$Ni. 
 We adopt $f_{\rm s}=2$ which corresponds to moderate mixing. 
The escape probability for gamma-rays through the metal layer 
is $\exp(-\tau_{\rm m})$,
where $\tau_{\rm m}$ is the optical depth of the metal layer, while 
the absorption probability in the mixing zone is described by the 
expression $\tau_1/(1+\tau_1)$, where $\tau_1$ is the optical depth 
of the inner zone.
In the mixing core gamma-rays have a finite probability of being 
absorbed by H-rich material, which is proportional to $1-f_{\rm mix}$.
The absorption probability for the outer hydrogen layer with 
the optical depth $\tau_2$ is defined as  $p_2=1-\exp(-\tau_2)$.

The energy deposited in the hydrogen is shared between ionization,
excitation and thermal energy with corresponding branching ratios
(\cite{Koz92}; \cite{Xu92}). The ionization of hydrogen is 
calculated in the approximation of three levels plus continuum. 
The photoionization from the second level by two-photon and 
continuum radiation was calculated assuming the continuum 
luminosity is equal to the deposited luminosity while the spectrum is
assumed to be dilute black-body with the temperature $T_{\rm c}$.
The electron temperature was set to be 5000~K.
Effects of radiation transfer in Balmer and Paschen continua are 
treated in the escape probability approximation (Chugai 1990).
 \begin{figure}
  \centering
  \vspace{9.5cm}
  \includegraphics{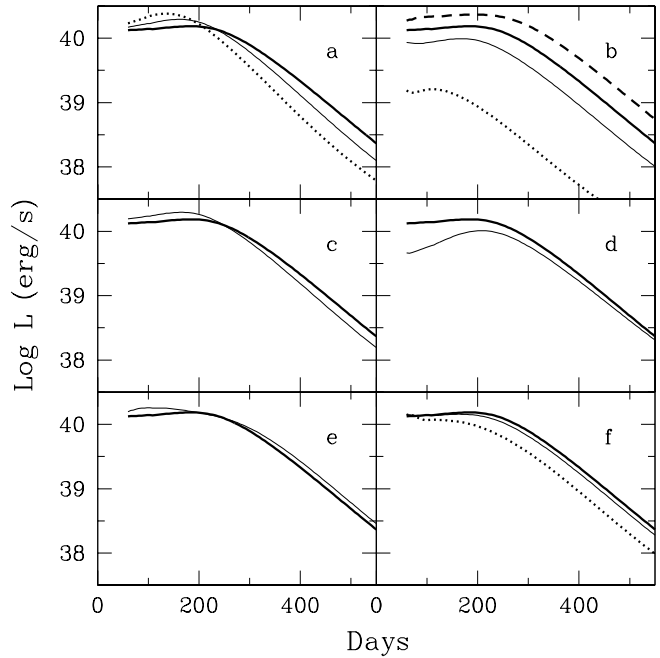}
  \caption[]{  The H$\alpha$ luminosity in the radioactive model. The 
  {\em thick line} shows the standard model, while
   {\em thin line} curves show models 
  {\em a, b, c, d, e, f}. Dotted lines in panels a,
 b and f correspond to models {\em a1, b1} and {\em f1} respectively. Additional model is shown in panel b (model {\em b2}, {\em dashed line}). Details of model parameters are presented in Table 3.
  }
  \label{hamod}
  \end{figure}

The sensitivity of the radioactive model to parameter variations 
(Table 3) is demonstrated in Fig.~\ref{hamod}. 
Each panel shows the effect of a single parameter variation
compared to a standard model, which is in fact the optimal  
model for the H$\alpha$ luminosity of SN~1987A. 
The saturation at the early epoch is a new feature compared to 
the previous model and this is 
 related to several factors: gamma-ray trapping in the inner zone,
collisional de-excitation of the third level and H$\alpha$ absorption 
in the Paschen continuum, especially in the mixing zone. In Fig.~\ref{hamod}a
we present models with lower mass (10 and $7~\rm M_{\odot}$) 
compared to the standard $14~\rm M_{\odot}$. A lower mass 
results in a high rate of the luminosity decay owing to a lower 
optical depth for gamma-rays. For moderate mass variation 
(factor $\leq 1.4$) the H$\alpha$ luminosity in the range $200-300$ d
shows less than 10\% deviation from the standard case. 
  With the rest of parameters fixed the H$\alpha$ luminosity is
proportional 
to the $^{56}$Ni mass (Fig. 6b) at the nebular epoch $t>200$ days. 
This property reflects the simple truth that to a good approximation 
the total ionization rate of hydrogen in the envelope is proportional 
to the total radiactive decay rate, {\rm i.e.} $^{56}$Ni mass. The 
proportionality, however, breaks down for $t<200$ days because 
of saturation effects related to H$\alpha$ collisional de-excitation 
and continuum absorption, which become important at early epoch.
The effect of factor 1.5 higher kinetic energy is similar to the effect of 
factor $\sim 1.2$ lower mass (Fig.~\ref{hamod}c), in accordance 
with the dependence of optical depth on both 
parameters $\tau_{\rm \gamma}\propto M^2/E$. 
A lower continuum temperature (Fig.~\ref{hamod}d) results in a
lower luminosity, especially, at the early epoch ($t<200$ d) 
since the ionization (and therefore recombination) rate becomes lower.
 A lower mixing core fraction 
(Fig.~\ref{hamod}e) results in the higher 
 H$\alpha$ luminosity since the total deposition in the outer H-rich 
 matter increases. 
The higher metal fraction in the mixing core
(Fig.~\ref{hamod}f), on the contrary, 
suppresses the late time H$\alpha$ luminosity consistent 
with the decreased deposition into mixed hydrogen material.
However, this effect is small if the variation of this parameter 
is within a factor 1.5.

The modeling shows that if ejecta parameters vary
 less than a factor 1.4 compared to those of SN~1987A, 
 the H$\alpha$ luminosity is then proportional 
 to the $^{56}$Ni mass (to within 10\% accuracy) at the late time nebular
 phase $200-400$ days. In this age range using the SN~1987A model as 
 a template we may estimate the $^{56}$Ni mass in other SNe~IIP 
 from the ratio of H$\alpha$ luminosity to that of SN~1987A. 
 It is worth noting here that despite 
 the large possible variation of the main sequence mass for SNe~IIP, 
 presumably say $10-25~\rm M_{\odot}$, the range in mass of the ejecta 
may actually be smaller since mass loss increases rapidly with the mass.   
\subsection {Results of $^{56}$Ni mass determination}
 The spectra utilized in this section
are based on data in the Asiago/ESO SN Catalogue. Additional spectra were kindly provided by R. Stathakis. A sample of the spectra is shown in Fig. 7 together with some line identifications. The available spectra
of our 
sample were corrected for reddening effects and the recovered integrated 
H$\alpha$ line fluxes are then translated to luminosities using adopted 
distances. In a straightforward approach we use the H$\alpha$ luminosity of 
SN~1987A as a template to recover the 
$^{56}$Ni mass from the H$\alpha$ luminosity of SNe~IIP 
assuming proportionality between the $^{56}$Ni mass 
and the H$\alpha$ luminosity in the range between $200-400$ days.
The resulting values of $^{56}$Ni mass are given in Table 4 
in the second column. 
\begin{figure}
\includegraphics[height=11.5cm,width=9cm]{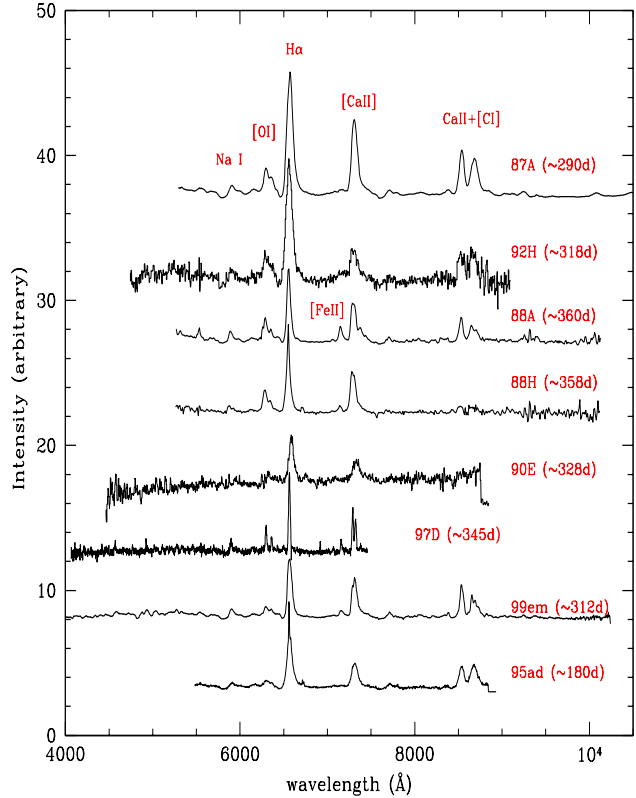}
\caption{ Sample of late time spectra of SNe IIP with the corresponding time since explosion. The spectra have been corrected for the recession velocities of their host galaxies. Some line identifications are also indicated.}
\end{figure}  

\begin{figure}
  \centering
  \vspace{9.8cm}
  \includegraphics{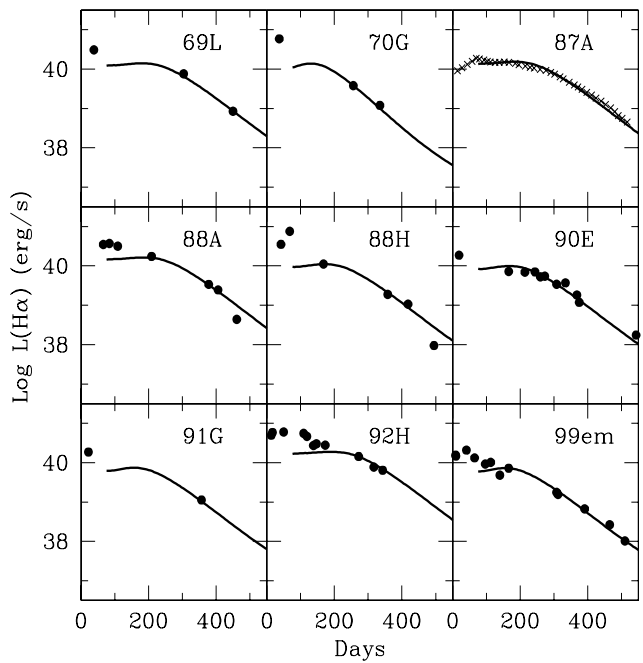}
  \caption[]{The H$\alpha$ luminosity evolution in different SN~IIP.
  Models ({\em solid lines}) are overplotted on the observational 
  data {\em dots} (sect. 4.2).
  }
  \label{hafit}
\end{figure}

Another approach has been to fit the H$\alpha$ light 
curves with the model of luminosity evolution. 
All the other parameters, 
except for $^{56}$Ni mass, are assumed to be the same as in the 
standard model for SN~1987A unless additional information indicates 
essentially different parameters. The plateau of SN 1970G is very brief 
($t_{\rm p}\approx 60$ d) which indicates a lower ejecta mass than that
of a typical SN IIP. We adopt $8~\rm M_{\odot}$ as the most appropriate value
for SN 1970G ejecta taking into account both the light curve and the rate
of the  H$\alpha$ flux decay.
The results are plotted for 9 SNe in Fig.~\ref{hafit}. Recovered values of 
$^{56}$Ni mass are given in Table 4 (third column).  
Note that the straightforward approach based upon the use of the 
H$\alpha$ light curve of SN~1987A (second column of Table 4) and the 
application of model fitting produce quite similar values with a maximal
 difference, apart from SN 1970G(SNP/L) and SN 1997D(treated bellow), of
 $\sim$13\% for SN 1988H. 

This deviation may be adopted as an uncertainty of the $^{56}$Ni mass
determination from the accurate data. Given the uncertainty of the 
observed value of the H$\alpha$ luminosity (at least 10\%) we are therefore
 able to derive a $^{56}$Ni mass from H$\alpha$ with an uncertainty of 
about 25\%.

\begin{figure}
  \centering
  \vspace{8cm}
  \includegraphics{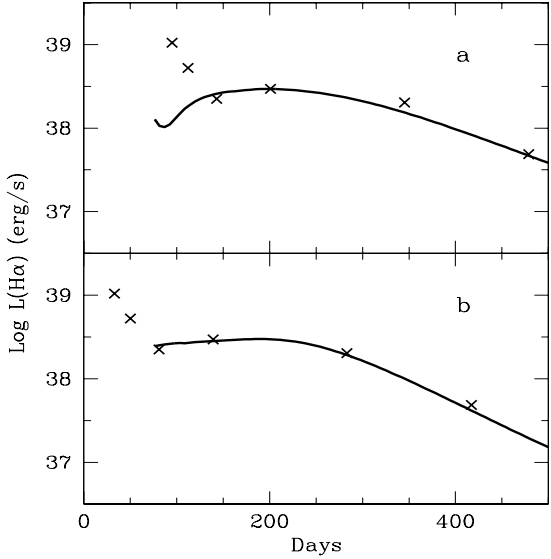}
  \caption[]{Modeling H$\alpha$ luminosity for SN~1997D.
  The upper panel shows the large age option, while lower panel 
  shows the small age option. Model results ({\em solid lines}) are
 overplotted on observational dots.
  }
  \label{97d}
  \end{figure}

The subluminous SN~1997D is a special case.
Originally, the age of the supernova at discovery 
 has been estimated to be  
 $t_{\rm d}=50$ days (Turatto et al. 1998), while 
 Zampieri et al. (2002) argue for an age almost twice as large 
$t_{\rm d}\approx95$ days. We consider both large and small age options. 
In the large age case we adopt 
$M=18~\rm M_{\odot}$, $E=9\times10^{50}$ erg 
(Zampieri et al. 2002), while in the small age option we adopt 
 $M=6~\rm M_{\odot}$ and $E=1\times10^{50}$ erg 
(\cite{Chug00}). A reasonable fit in the first case 
(Fig.~\ref{97d}a) is found for the $^{56}$Ni mass
of $0.011~\rm M_{\odot}$ with a mixed core fraction $f_{\rm mix}=0.05$, which 
corresponds to the mixing core having $0.9~\rm M_{\odot}$ and a core metal fraction
$f_{\rm m}=0.99$, {\rm i.e.} with the metal core practically 
devoid of hydrogen.
The model also requires a low continuum temperature in the ultraviolet 
$T_{\rm c}=4500$~K. For the small age choice the fit 
(Fig.~\ref{97d}b) corresponds to  
$M(^{56}\rm{Ni})=0.0024~\rm M_{\odot}$. In this case $f_{\rm mix}=0.05$, 
$f_{\rm m}=0.3$ (a significant amount of mixed hydrogen) and 
$T_{\rm c}=5500$~K.

\begin{figure}
\includegraphics[height=9.5cm,width=9cm]{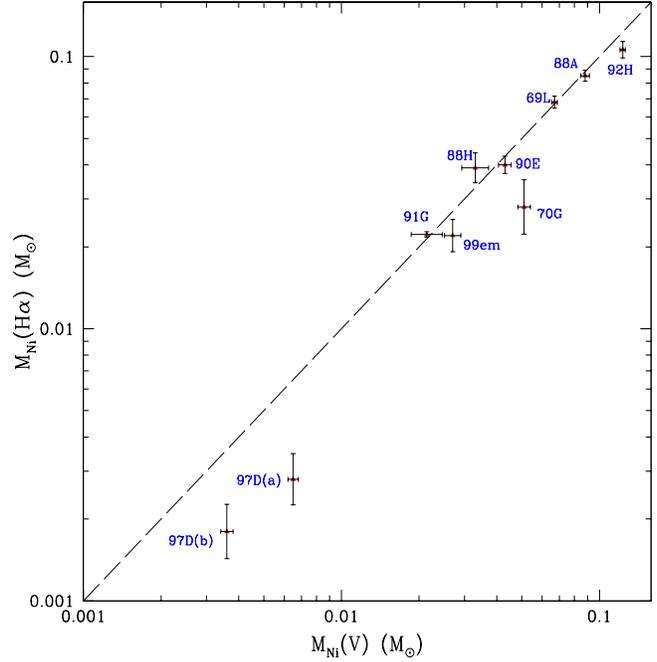}
\caption{ The correlation between the $^{56}$Ni mass derived from  
the H$\alpha$ luminosity, $M_{\rm Ni}(\mbox{H}\alpha)$, and that 
 derived from the tail $M_V$ magnitude, $M_ {\rm Ni}(V)$.
The dashed line has a slope of unity. Clear deviation is seen for SN 1970G(SNP/L) and for both scenarios of the faint event SN 1997D.}
\label{phsp}
\end{figure}

The $^{56}$Ni mass derived from  
the H$\alpha$ luminosity adopting SN 1987A as template, $M_{\rm Ni}(\mbox{H}\alpha)$, 
is plotted in Fig.~\ref{phsp} versus photometric $^{56}$Ni mass,
$M_ {\rm Ni}(V)$,  derived from the tail $M_V$ magnitude (Table 2). 
Both sets of values agree within 20\% . This consistency supports 
the proposition that the H$\alpha$ luminosity may be 
 a good indicator of $^{56}$Ni mass in SNe~IIP unless we are dealing with 
extreme cases such as SN~1970G (SNP/L) or underluminous cases such as SN~1997D.
\subsection{Application of H$\alpha$ to $^{56}$Ni diagnostics}
\begin{table}
\begin{minipage}{80mm}
 \caption{Recovered mass of $^{56}$Ni using SN 1987A as template (second column), and from late time L(H$\alpha$) modeling ``$M_{\rm Ni}^{m}(H\alpha)$''.} 
\bigskip
\centering
\begin{tabular}{c c c c }
\hline \hline
SN  & $M_{\rm Ni}(\rm H\alpha)$& $M_{\rm Ni}^{m}(\rm H\alpha)$ \\
name &$(M_{\odot})$&$(M_{\odot})$  \\
\hline
1969L &0.068($\pm0.004$)& 0.065\\
1970G &0.029($\pm0.011$)&0.035\\
1988A &0.085($\pm0.004$)&0.083 \\
1988H &0.039($\pm0.005$)&0.045 \\
1990E &0.04($\pm0.003$)&0.038 \\
1991G &0.022($\pm0.0005$)&0.025 \\
1992H &0.106($\pm0.008$)&0.105 \\
1997D(a) &0.0028($\pm0.0006$)&0.011 \\
1997D(b) &0.0018($\pm0.0004$)&0.0024 \\
1999em&0.022($\pm0.003$)&0.024 \\
\hline \hline 
\end{tabular} 
\end{minipage}\\
\label{tnisp} 
\hspace{-7truecm}\emph{\rm }

\normalsize
\end{table}
In the absence of the late time photometry but with available spectra 
of SNe~IIP at the nebular epoch we may use the H$\alpha$ luminosity
to estimate $^{56}$Ni mass ejected by supernova provided the explosion time
 is known. We demonstrate this approach 
to SN 1995ad, SN 1995V and SN 1995W (all type IIP) 
 for which late photometric data 
are not available. Distances are computed using the corrected 
recession velocities (reported in ``LEDA'' data base assuming 
$H_{0} =$70 km s$^{-1}$Mpc$^{-1}$).

\bf{SN 1995ad:} \rm This SN was discovered on Sep. 28.8 UT in NGC 2139 by 
R. Evans (\cite{IA39}). Its Sep. 29.3 UT spectrum, obtained by 
S. Benetti, displays a hot continuum with temperature $T_{bb}$ $\sim$ 
13000 K and a broad H$\alpha$ emission ($FWHM=$11000 km s$^{-1}$). 
 These characteristics combined with the fact that nothing was 
 seen at the position of the SN in the Aug. 25 image (IAUC 6239) 
 provide constraints on the explosion time. The assumed distance and 
 reddening are respectively $D=23.52$ Mpc and $A_V =0.112$~ mag 
 (NED; Schlegel et al. 1998) while the galaxy inclination is $40.9^{\circ}$. 
The spectrum we use here is taken 1996 March. 24, about 180 days 
since explosion. The derived H$\alpha$ luminosity,
 $\sim 1.06\times10^{40}$ erg s$^{-1}$, is then compared to 
that of SN 1987A at a similar epoch to recover an ejected mass 
$\approx 0.056~\odot M_{\odot}$. 

\bf{SN 1995V:} \rm discovered in NGC 1087 by R. Evans (\cite{IA97}). 
The galaxy inclination is 33.2$^{\circ}$.
The adopted distance and 
reddening are, accordingly, $D=20.61$ Mpc and $A_V =1.37$~ mag,
while the explosion time is assumed to be 1995 July. 25 (\cite{Fass98}).
The available late spectrum was taken around day 409 since explosion. 
Comparing the recovered H$\alpha$ luminosity, 
$\sim 1.33\times10^{39}$ erg s$^{-1}$, with that of SN 1987A 
leads to the estimate of $M(^{56}\mbox{Ni})\approx 0.046~\rm M_{\odot}$.

\bf{SN 1995W:} \rm discovered on Aug. 5.65 UT in NGC 7650 by A. Williams and R. Martin (\cite{IA06}). The 
spectrum taken on Aug 17.29 UT displayed features of SN~IIP 
around one month (IAUC 6206). The distance is $D=44.81$ Mpc, 
and the reddening is $A_V =0.044$~ mag (NED; Schlegel et al. 1998) and the host galaxy inclination is 47.4$^{\circ}$. The late spectrum was taken around day 300 (1996 May. 12). 
The H$\alpha$ luminosity, 
$\sim 4.79\times10^{39}$ erg s$^{-1}$, when compared to that of SN 1987A 
at similar epoch yields an amount of $^{56}$Ni mass of  
 $M(^{56}\mbox{Ni})\approx 0.048~\rm M_{\odot}$.   
Note that the adopted reddening values for SN 1995ad and SN 1995W 
 do not include the host galaxy reddening. For these events the inclination of their host galaxies may provide further indication of uncertainty in the extinction since large extinctions might be expected for highly inclined galaxies if the SN occurred towards the far side. Although no sign of high extinction from interstellar lines was reported in early spectra of these two objects, the derived amounts of $M(^{56}\rm Ni)$ might be taken as lower limits bearing in mind the significant inclination of their host galaxies.
\section{Discussion and conclusion}
The primary goal of this paper was to demonstrate that 
the H$\alpha$ luminosity at the nebular epoch may be an 
indicator of the $^{56}$Ni mass in SNe~IIP ejecta. To explore this 
idea we selected a sample of well observed SNe~IIP, and 
using $V$ magnitudes on the radioactive tail 
we derived in a standard way the photometric mass of $^{56}$Ni mass.
With our sample we confirmed using a slightly different approach 
the correlation between $^{56}$Ni mass 
and absolute magnitude $M_V$ of the plateau reported recently by 
Hamuy (2003). 
We then applied a two-zone model of the H$\alpha$ luminosity in 
SN~IIP to explore the sensitivity of the H$\alpha$ behaviour to 
variation of model parameters. We found that if mass, energy and 
mixing conditions do not vary strongly among SNe~IIP  
(less than factor 1.4) then with an accuracy better than 10\% 
H$\alpha$ luminosity is proportional to $^{56}$Ni mass during 
the $200-400$ days after explosion.
H$\alpha$ luminosities were then used to derive $^{56}$Ni masses.
This was done employing two approaches: first, 
using the H$\alpha$ light curve in SN~1987A as template and, second, 
applying the model computations. Both approaches agree  
within 15\% unless we are dealing with extreme cases such as SN~1970G
(type IIP/L) and underluminous SN~1997D.
In both these cases we should possess additional information 
about ejecta mass and energy to derive the $^{56}$Ni mass from H$\alpha$.
The $^{56}$Ni mass values derived from the photometry and 
H$\alpha$ luminosity agree  within 20\%, which thus gives us 
confidence that H$\alpha$ is a good indicator of the amount of 
$^{56}$Ni in SNe~IIP. Simultaneously, this consistency suggests that 
parameters of SNe~IIP (mass, energy and mixing) are not very 
different. In fact this is consistent with the uniformity of 
plateau luminosities and plateau lengths of SNe~IIP.

We applied the  method of $^{56}$Ni mass estimation from H$\alpha$ 
for three SNe~IIP (SN~1995ad, SN~1995V and SN~1995W) 
without photometry at the nebular phase and 
derived $^{56}$Ni masses of $\approx 0.056~\rm M_{\odot}$, 
$0.046~\rm M_{\odot}$, $0.048~\rm M_{\odot}$, quite reasonable values, 
although they could be underestimates, since the host galaxy extinction 
(for SN 1995ad \& 1995W) has not been taken into account. Nevertheless, this 
is a good demonstration of the possibility of the method.
Generally, the approach based upon H$\alpha$ 
 may be indispensable in cases, when 
the photometry at the nebular epoch is absent, or when there is 
a problem with subtraction of stellar background (SN~IIP in the bulge, 
or in high redshift galaxies).

An interesting by-product of the analysis of the SNe~IIP 
sample is the demonstrated correlation between $^{56}$Ni mass and the 
steepness parameter ($S$) introduced to measure the light curve decay 
rate at the inflection point. The correlation is such that the
steeper the decline at the inflection point the lower is the 
$^{56}$Ni mass. Thus radiative diffusion times and $^{56}$Ni masses are 
linked. How an increased radioactive energy input leads to a higher effective
 opacity will require elaboration by hydrodynamical modeling.
This correlation, if confirmed, will provide us with a distance and 
extinction independent way to determine the amount of ejected $^{56}$Ni.

Finally we note that in Fig. 10 the clustering of
the points around two values of $^{56}$Ni mass viz. 0.005 and 0.05 $\rm M_{\odot}$ may result 
from poor statistical sampling. On the other hand it may be a hint that a mechanism such as fall-back is an important one in the evolution of the low-mass group.
\begin{acknowledgements}
We thank Raylee Stathakis for providing us with some
unpublished spectra in the AAO Archive which were useful for measurements of H$\alpha$ luminosities. A. Elmhamdi thanks A. Pastorello
for providing some data for SN 1999eu.  Research by N. N. Chugai 
was supported by grant ``RFBR 01-02-16295'' of Russian Academy of Sciences. 
\end{acknowledgements}

\end{document}